\documentstyle[aps,prb]{revtex}
\begin{document}
\begin{center}
{\large\bf Multiphonon anharmonic decay of a quantum mode}

\vspace{2ex}
V.Hizhnyakov

\vspace{1ex}
{\it Institute of Theoretical Physics, University of Tartu,
T\"{a}he 4, 51014 Tartu, Estonia}\\
{\it Institute of Physics, University of Tartu,
Riia 142, 51014 Tartu, Estonia} 

\end{center}

\begin{abstract}
A nonperturbative theory of multiphonon anharmonic transitions between
energy levels of a local mode is presented. It is shown that
the rate of transitions rearranges near the critical level number $n_{cr}$:
at smaller $n$ the process slows down, while at larger
$n$ it accelerates in time, causing a jump-like loss of  energy
followed by the generation of phonon bursts. Depending on parameters, phonons 
are emitted in pairs, triplets etc. 
\end{abstract}
\vspace{5mm}

\vspace{10mm}
The nonlinear dynamics of strong vibrational excitations in crystals has
attracted remarkable attention during the last years. The opportunities of
applying novel numerical techniques play an essential role in the
investigation of nonlinear discrete systems, allowing for the observation of
different  phenomena. One of the effects,  the existence of
localized vibrations in a pure anharmonic lattice (so-called intrinsic local
modes (ILM) or self-localized solitons (SLS))
\cite{ovchinn,kosevich,dolgov,sivtak,flach}, is of particular interest, 
as it links
local lattice dynamics to the physics of solitons and underlines, in general,
the importance of strongly excited local modes. So far the research of
strong
anharmonic effects in lattice dynamics as well as molecular dynamics
simulations of a nonlinear lattice have been mainly carried out
within the framework of classical mechanics. However, quantum effects are 
important
here as they lead to extra mechanisms in the decay of vibrational
excitations. This holds especially for the modes with frequencies above
the phonon spectrum  because they, being singly excited, do not decay at
all in the classical limit -- it is because 
all their harmonics are out of 
resonance with the linear spectrum which makes it impossible to transfer 
energy to phonons \cite{flach}.
Quantum decay of the ILMs
has been discussed by Ovchinnikov \cite{ovchinn} in the framework
of standard perturbation
theory. However, this theory may not be applicable
here because ILMs exist only when their amplitudes (energy) are large enough;
therefore, one cannot assume that anharmonic interactions are weak.
To understand how relaxation proceeds when the
energy of the mode is large, one should develop a nonperturbative theory
of multiphonon transitions between the quantum levels
due to anharmonic interaction with the phonon continuum.

Recently, in \cite{hizhrev,zhizh} we proposed 
a theory  of anharmonic damping of a strongly excited mode, 
supposing that the latter can be considered classically
(phonons in this theory are described quantum-mechanically).
The idea of Ref. \cite{hizhrev,zhizh} was to consider the effect of the
strongly excited mode upon the phonon operators. 
The goal was to find a transformation of these
operators in time, and, from that, to obtain the probability of creation of a
phonon. Adding all the created phonons allows one to
find the energy loss in the mode.

The realization of the method in the case of
quadratic interaction  of the mode with respect to phonon coordinate 
operators $\hat{x}_i$ is straightforward:
the phonon Hamiltonian with the anharmonic interaction
$\sum_{ii'}V_{lii'}(x_l(t))\hat{x}_i \hat{x}_{i'}$,
where $x_l(t) =
A \cos{(\omega_lt)}$ is a classical time-dependent coordinate of the local
mode, can be diagonalized exactly for an arbitrary
amplitude $A$ of the mode. This is achieved by the transformation
$\hat{b}_j (t) = \sum_i[\mu_{ij}(t) \hat{a}_i + \nu_{ij}(t) \hat{a}^+_i]$
where $\hat{a}_i$ and $\hat{a}^+_{i'}$ are initial destruction and creation
operators of phonons
(for explicit expressions for $\mu_{ij}$ and $\nu_{ij}$, see  \cite{hizhrev}).
The initial zero-point state, $\vert 0\rangle$,
is not the zeroth state of the operators $\hat{b}_j$.
This means that phonons are generated in a lattice.
The number of created phonons $j$ equals
$ n_j = \sum_{i} \vert \nu_{ij}\vert^2$. For
times $t \gg \omega_i^{-1}$ the transformation
simplifies  \cite{zhizh}: $\mu_{ij}, \nu_{ij} \sim \delta_{ij}$ and the rate
of energy relaxation becomes
\begin{equation}
-\dot{E}_l \simeq \sum_i \hbar \omega_i d |\nu_i|^2/dt.
\end{equation}

In \cite{hizhrev}, it was found that for a large energy mode, the
relaxation rate diminishes with energy: $-\dot{E}_l \sim A^{-2}$,
i.e. relaxation
accelerates with time. This result contradicts standard perturbation theory.
Yet, the interpretation of this surprising result is straightforward -
the decay is of purely
quantum mechanical origin and, therefore,
it should disappear in the classical limit
$E_l \rightarrow \infty$. It was also shown that the acceleration of the
process continues down to an energy $E_{cr}$ (which typically corresponds to
$E_{l}\sim 10 - 30$ vibrational quanta, or $A\sim 0.2\AA - 0.5\AA$ ) when
the rate becomes high (of the order of $\hbar \omega_l^2$)
and a sharp relaxation jump
takes place accompanied by generation of a burst of phonons.
The latter are emitted in pairs.
After the first jump, several other jumps may occur. For small energies
relaxation slows down exponentially in agreement with perturbation theory.

The theory of Ref.  \cite{hizhrev,zhizh} considers the relaxation of
the mode as a continuous
process where the energy of a classical subsystem is diminishing. In this
communication, the effects of the quantum origin of the states of the mode 
and of the transitions between its levels will be considered. Here we restrict
ourselves to quadratic terms in the expansion of the potential
energy with respect to phonon operators: $\hat{U} = U_0(\hat{x}_l) +
 \sum_i U_{1i}(\hat{x}_l)\hat{x}_i + 
\sum_{ii'} U_{2ii'}(\hat{x}_l)\hat{x}_i \hat{x}_{i'}/2$; the
potential of the local mode, $U_0(\hat{x}_l)$, is supposed to be 
anharmonic, and so are the factors $U_{1i}(\hat{x}_l)$
and $U_{2ii'}(\hat{x}_l)$: they depend on the
displacement operator $\hat{x}_l$ due to higher order anharmonicities. 
To clarify the range of the 
applicability of the approximation, we observe that all anharmonic 
terms become comparable to each other for shifts of an atomic distance of
typically $1 \AA$ \cite{klinger}.  E.g. the harmonic
$V_l^{(2)}x_l^2$ 
and anharmonic
$(V_l^{(3)}/3 + V_l^{(4)}x_l/12)x_l^3$  terms
in the expansion of $U_0(x_l)$
are of the same order of magnitude
for $x_l \sim 1 \AA$, while the terms
$V_{ii'}^{(2)}x_i x_{i'}$ and $(V_{lii'}^{(3)} +
V_{lii'}^{(4)} x_l/2) x_l x_i x_{i'}$
in the expansion of $U_{2ii'}(x_l)x_i x{i'}$ 
are of the same order of magnitude for $x_l \sim 0.3 \AA$. 
Below we take into account these terms while we consider just local modes with 
the amplitude $A \sim 0.3 \AA$ or less. For such amplitudes the terms 
$\sim x_i x_{i'} x_{i"}$ are already much  smaller and are neglected.

Within this approximation,
the interaction of the local mode with phonons  has the form
\begin{equation}
\hat{H}_{int} =\sum_i V_{lli}(\hat{n}) \hat{x}_i +
\frac{1}{2}\sum_{ii'}[V_{lii'}(\hat{n})\hat{a}_l+ H.c.]
\hat{x}_i \hat{x}_{i'},
\end{equation}
where $\hat{n}= \hat{a}_l^+ \hat{a}_l$, $\hat{a}^+_l$ and $\hat{a}_l$ are
the creation and destruction operators;
terms $\sim \hat{a}^{+^k}_l$ as well as $\sim \hat{a}_l^k$, $k=2,3,...$
are neglected; they describe transitions between the levels 
$n$ and $n-k$ and could be considered analogously. 
Below the operators  $V_{lli}(\hat{n})$ and $V_{lii'}(\hat{n})$ 
stemming from $U_{1i}$ and $U_{2ii'}$ respectively,
are applied only to the initial number  state $|n\rangle$ of the local mode. 
Therefore the $V(\hat{n})-$operators can be replaced by 
$C$-functions of $n$. In this approximation, 
using shifted phonon operators $\hat{x}'_i=
\hat{x}_i+x_{0i}$, with $x_{0i}= V_{lli}\omega_i^{-2}$,  
the interaction part of the hamiltonian reads
$$
\hat{H}'_{int} =\sum_i\hat{V}_{li} \hat{x}'_i +
\frac{1}{2} \sum_{ii'}V_{lii'} (\hat{a}^+_l + \hat{a}_l)
\hat{x}'_i \hat{x}'_{i'},
$$
where $\hat{V}_{li} = \bar{V}_{li} (\hat{a}^+_l + \hat{a}_l)
+ V_{lli}(\hat{n}-n)$,
$\bar{V}_{li}= - \sum_{i'} V_{lii'}V_{lli}\omega_i^{-2}$
(all $V-$factors depend on $n$).

Let us present the operators of the phonon coordinates in the form
$\hat{x}'_i(t) = (\hbar/2\omega_i)[\hat{g}_i(t) \hat{a}_i + H.c.]$.
Here the operators $\hat{g}_i(t)$ account for the time dependence of
$\hat{x}'_i(t)$. This dependence is determined by the equations of 
motion
\begin{equation}
\ddot{\hat{g}}_i+\omega_i^2 \hat{g}_i =
- \omega_i(\hat{V}_{li} + 
\sum_{i'}V_{lii'} \hat{g}_{i'}(\hat{a}_l + \hat{a}_l^+))
\end{equation}
and by the initial conditions
$\hat{g}_i(t) = e^{-i\omega_it+i\phi_i}$,$\,\,t \rightarrow 0$,
$\phi_i$ is the random phase 
(the final expression should be averaged over $\phi_i$).
The operators $\hat{g}_i$ do not commute with
$\hat{a}_l + \hat{a}^+_l$; the order of the operators in 
right-hand side of Eq. (3)
depends on whether one considers the operators $\hat{g}_i$ or the operators
$\hat{g}_i^+$; the opposite order describes $\hat{g}^+_i$. 
The integral form of the equations (3) is the following:
$$
\hat{g}_i(t) =e^{-i\omega_it+i\phi_i} +
\int_0^t dt'\sin{(\omega_i(t-t'))}[\hat{V}_{li}(t')
+ \sum_{i'} (\bar{e}_i V_l \bar{e}_{i'})
\hat{g}_{i'}(t') (\hat{a}_l(t') + \hat{a}^+_l(t'))].
$$
Here $(\bar{e}_{i} V_{l} \bar{e}_{i'}) \equiv V_{lii'}
= \sum_{mm'}\bar{e}_{im} V_{lmm'} \bar{e}_{i'm'}$, $m$ is the
index of Cartesian coordinates of the atoms in the lattice,
$\bar{e}_{im}= e_{im}\omega_i^{-1/2}$, $e_{im}$ are projections
of the reduced displacements $x_m$ to the normal coordinates. 

We assume that relaxation of the mode is slow 
as compared to its  frequency. Then describing
the decay of the $n$-th level caused by the $n \rightarrow n-1$
transition, one can neglect all other levels of the mode.
This means that one can use the rotating wave 
approximation when  only terms with creation and destruction operators, 
being arranged alternatingly
in the iteration series for the 
$\hat{g}_i$, are taken into account.
Let us consider in this approximation the large time asymptotics of the 
two-time correlation function
$\langle n|\hat{x}'_i (t+\tau)\hat{x}'_i(t)|n\rangle$ for
$t\gg \tau \sim \omega_i^{-1}$. Here $|n\rangle$ denotes the $n$th
state of the mode and the ground state of phonons. 
Only the matrix elements $\langle n|\hat{g}_i|n\rangle$
and $\langle n|\hat{g}_i|n -1\rangle$ contribute to this asymptotics
($|n -1\rangle \equiv n^{-1/2} \hat{a}_l |n\rangle$).
The terms $\sim V_{lli}(\hat{n}-n)$  of the expression for $\hat{g}_i(t)$
give zero; the terms $\sim \bar{V}_{li}
(\hat{a}_l + \hat{a}_l^+)$ of this expression
oscillate fast and, therefore, give small contribution which can be 
neglected (note, that the basic time-dependence of the 
operators $\hat{a}_l$ and $\hat{a}^+_l$ is  given by the factors 
$e^{-i\omega_l t}$ and $e^{i\omega_l t}$  respectively). In this approximation
$\langle n|\hat{x}'_i (t+\tau)\hat{x}'_i(t)|n\rangle \approx
(\hbar/2 \omega_i)[|\mu_i(t)|^2 e^{-i\omega_i \tau} +
|\nu_i(t)|^2 e^{i\omega_i \tau}],\,\,t\gg \tau \sim \omega_i^{-1}$, where
$$
\mu_i(t) \approx e^{i\phi_i} 
-(i/2) \int_0^t dt' \sum_{i'}(\bar{e}_i V_l \bar{e}_{i'})
e^{i (\omega_i + \omega_{i'}) t'}\nu_{i'}(t') 
\langle n-1| \hat{a}_l(t')|n\rangle,
$$
$$
\nu_i(t) \approx (i/2) \int_{0}^{t} dt'
\sum_{i'}(\bar{e}_i V_l \bar{e}_{i'})
e^{-i (\omega_i + \omega_{i'}) t'}\mu_{i'}(t')
\langle n| \hat{a}^+_l(t')|n-1\rangle
$$
($\mu_i \sim \langle n|\hat{g}_i|n\rangle$, 
$\nu_i \sim \langle n|\hat{g}_i|n-1\rangle$;
here it was taken into account that 
$\langle n|\hat{a}_l^+(t')|n-1\rangle
\langle n-1|\hat{a}_l(t)|n\rangle \cong 
\langle n|\hat{a}_l^+(t')\hat{a}_l(t) |n\rangle $).  
This asymptotics corresponds to the following transformation 
of phonon operators :
$\hat{b}_i \simeq \mu_i\hat{a}_i + \nu_i^* \hat{a}_i^+$.
To find this transformation explicitly we  present the above
expression for $\nu_i(t)$ in the form
\begin{equation}
\nu_i(t) \approx (i/2) \int_0^t dt' e^{-i\omega_i t'} 
(\bar{e}_i V_l D(t'))\langle n|\hat{a}^+_l(t') |n-1\rangle.
\end{equation}
Here $D(t) = \sum_i e^{-i \omega_i t} \bar{e}_i \mu_i (t)$ 
satisfies the equation
\begin{equation}
D(t)=D_{0}(t)
+\int_{0}^{\infty}dt'\int_{0}^{\infty}dt"G(t-t')
V_l G^*(t'-t")V_l F(t' - t")D(t"),
\end{equation}
$D_{0}(t)=\sum_{i}\bar{e}_{i}e^{-i\omega_{i}t+i\phi _{i}}$,
$G_{mm'}(t)
=-(i/2)\Theta (t)\sum_{i}\bar{e}_{im}\bar{e}_{im'}e^{-i\omega_i t}$
is the  Green's function of phonons, $\Theta (t)$ the Heaviside step-function,
and
\begin{equation}
F(t)\simeq  \langle n|\hat{a}_l^+\hat{a}_l(t)|n\rangle.
\end{equation}

Correlation functions of this type are well known in the theory
of optical transitions (see, e.g.  \cite{Lax}): $F(t) =
ne^{-i\bar{\omega}_l t+f(t)}$, where
$f(t)= - \Gamma t +\sum_i(V_{lli}^{2}/2\hbar\omega_i^3)
(e^{i\omega_it}-1)$; the sum 
accounts for phonon transitions caused by a
shift in their equilibrium positions in connection with the
$n \rightarrow n-1$ transition in the mode, $\Gamma= 
-\dot{E}_l/\hbar\bar{\omega}_l$ is the decay constant, which should be
determined self-consistently (if the relaxation is slow as compared to 
characteristic rates of the phonon Green's functions, then
one can neglect the term $\Gamma t$ in this expression);
$\hbar \bar{\omega}_l= \hbar\omega_l-
\sum_iV_{lli}^{2}/2\omega_i^2$
is the renormalized level spacing of the local mode.
Equation (5) can be easily solved by means of a half-axis Fourier transform.
One gets
\begin{equation}
D(\omega) = R(\omega)D_0(\omega);\,\,\,\,
R(\omega) = [I - G(\omega)v\bar{G}^*(\omega_l-\omega)v]^{-1}D_0(\omega)
\end{equation}
($v=V_l \sqrt{n}$). Here the functions $D_0(\omega)$,
$G(\omega)$ and  $\bar{G}^*(\omega)$
are the half-axis Fourier transforms of the functions
$D_0(t)$, $G(t)$ and $G^*(t) e^{f(t)}$.

Inserting Eqs. (4) - (7) into (1) and averaging over $\phi_i$ 
(leading to the replacement of $e^{i(\phi_i-\phi_{i'})}$ by $\delta_{ii'}$), 
one gets the following formula for the damping rate of the $n$th level
of the mode:
\begin{equation}
\Gamma(n) \simeq \frac{1}{2\pi } \int_0^{\infty}d\omega
Sp[R(\omega )v\rho(\omega)R^*(\omega)v\bar{\rho}(\bar{\omega}_l-\omega)].
\end{equation}
Here
$\rho(\omega )=ImG(\omega)$ and $\bar{\rho}(\omega)=Im\bar{G}(\omega)$
are the usual and renormalized phonon density functions, respectively.
The damping rate of the energy is $\dot{E}_{l,n} = 
\hbar\bar{\omega}_{ln}\Gamma(n)$.
A comparison of this result with an analogous formula, obtained
within the classical description of the local mode, \cite{hizhrev}
shows their coincidence
in the limit $f(t)=0$ ($\bar{G}(\omega) = G(\omega)$). 
Such a difference should be expected: the function
$f(t)$ accounts for phonon transitions, assisting the destruction
of a quantum of the local mode, thus being of quantum mechanical origin.

For moderately large $n$,$\,$ 
$\Gamma(n) \sim v(n)^{-1}$,
i.e. the relaxation accelerates with time, which is analogous
to $-\dot{E}_l \sim A^{-2}$ obtained
in the classical description of 
the local mode \cite{hizhrev}. 
The acceleration continues down to the level $n_{cr}$,  
when the resolvent in $R(\omega )$ approaches zero 
(for $\omega \approx \bar{\omega}_{l}/2$ and in some cases for other
$\omega $). One can show \cite{hizhrev} that near such a level, 
$\Gamma(n) \sim |n-n_{cr}|^{-1}\sim |t-t_{cr}|^{-1/2}$, i.e. a 
sharp relaxation 
jump takes place accompanied by the generation of a burst of phonons 
(time $t_{cr}$ corresponds to $n=n_{cr}$). After a first jump, several
other jumps may occur. For small $n$, formula (8) coincides with the result
given by Fermi's Golden Rule (i.e. the standard perturbation theory).
In the classical limit $\hbar \rightarrow 0$, the
decay of the mode
energy vanishes for phonons in the ground state.

The renormalized phonon density function $\bar{\rho}(\omega)$ is 
nonzero also for $\omega$
larger than the maximal phonon frequency $\omega_M$
(due to phonon-assisted
transitions described by the function $f(t)$),
while $\rho(\omega)$ is nonzero only in the one-phonon
region $0< \omega < \omega_M$.  Therefore formula (8)
accounts not only for two-phonon decay of the
local mode but also for its three- and more-phonon decay
(the corresponding formula 
for a classical local mode \cite{hizhrev} 
accounts for two-phonon decay only). The
linear and quadratic interactions considered above
are commonly present, enhancing or weakening each
other, as follows from the value of the resolvent $R(\omega)$ in (8). As
a result, three- and more-phonon processes may also cause relaxation
jumps, e.g. if $\omega_l \geq 2\omega_M$ and two-phonon processes are 
forbidden by the energy conservation law. If $\omega_l \gg 2\omega_M$, 
the consideration of linear and quadratic terms with respect to
phonon operators in the interaction hamiltonian is not sufficient; one needs
to include higher order terms (see, in this connection,
ref.\cite{zhizh}, where these terms have been
considered  within the classical description of the local mode).

To illustrate the effect of
three- and more-phonon  processes according to the quantum
formula (8), we performed a model calculation
of the relaxation rates
of a ${\rm Xe}_2^*$-molecule in solid ${\rm Xe}$. Note that the relaxation of 
this system was  recently studied experimentally in
\cite{xenon} where the
relaxation jump was found to take place near the level $n = 22$.
The level spacing in a ${\rm Xe}^*_2$-molecule is equal to
$\hbar \omega_M (68-n)/24$. Two-phonon relaxation
is allowed for upper levels until $n=20$.  Lower levels
can decay with emission of no less than three phonons  (their
spacing exceeds $2\hbar \omega_M$).
Only the symmetrized displacement
of two nearest atoms along the ${\rm Xe}_2^*$-molecule contributes
significantly to the anharmonic interactions of this molecule
with the environment (in this approximation,
all matrices in (8) are scalars); the main contribution to this
displacement comes from the longitudinal phonons \cite{Anselm}.
The corresponding density function is well  approximated
by the cap-type shape $\rho(x) = 8x\sqrt{1-x^2}$, $x=\omega/\omega_M$
\cite{Lurie}.
Assuming $f(t)=-\Gamma t + \beta (G(t)-G(0))$,
we calculated the $\Gamma(n)$-function for different values of the parameter
$\beta \sim V_{llm}$ (determining the shifts in the equilibrium positions 
of the phonon coordinates with the
$n \rightarrow n-1$ transition of the local mode) 
and with the maximum at $n=22$.
The results for 
$\gamma(n)=\Gamma(n) \omega_{ln}/\omega_M$ are presented in Fig.1.
One sees that not only the two-phonon decay but also 
the three phonon decay is strongly enhanced in the vicinity of
the critical level $n=22$. An estimation \cite{xenon}  
gives for ${\rm Xe}_2^*$-molecule in ${\rm Xe}$ crystal $\beta\approx 0.2$.

Above the case $T=0$ was considered. 
Finite temperature effects are straightforward to take into account:
in Eq. (8), there would appear an additional factor $1+2n(\omega)$
under the integral and the factor $(e^{i\omega_it}-1)$
in $f(t)$ should be replaced by the temperature-dependent factor
$(n(\omega)+1)(e^{i\omega_it}-1) + n(\omega)(e^{-i\omega_it}-1)$ (here
$n(\omega)= 1/(e^{\hbar \omega/kT}-1)$). The
formulae obtained are applicable generally for 
nonadiabatic transitions in the case of
linear diagonal and quadratic nondiagonal interactions of the system with
the phonon continuum.

In conclusion, we have presented a nonperturbative theory of multiphonon
anharmonic quantum transitions between levels of a local mode taking into
account linear and quadratic terms with respect to phonon operators. 
We showed that for high levels, transitions accelerate
in time which is in contradiction with the standard  perturbation theory.
This continues down to the level $n_{cr}$, when the rate of the process becomes
large - of the order of $\omega_M^{-1}$, and a sharp relaxation jump
takes place, accompanied by the generation of a burst of phonons. Depending 
on the level spacing, phonons may be emitted in pairs or triplets, quartets
and etc. After the first jump, several other jumps may occur. Only
for low levels  standard perturbation theory and Fermi's Golden Rule are
applicable. This theory can be applied generally for the nonperturbative
description of nonadiabatic transitions
caused by linear diagonal and quadratic nondiagonal interactions of the
system with the phonon continuum.

The author is grateful to Dr. M. Selg for his assistance concerning the
enhancement of the three-phonon relaxation.
This research was supported by Estonian Science Foundation,
Grant No.~2274 and by the DAAD Grant HSPS.

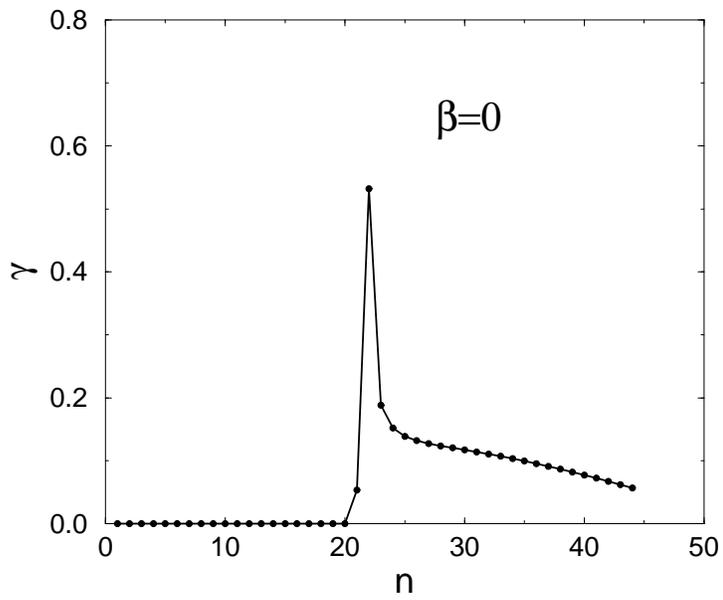
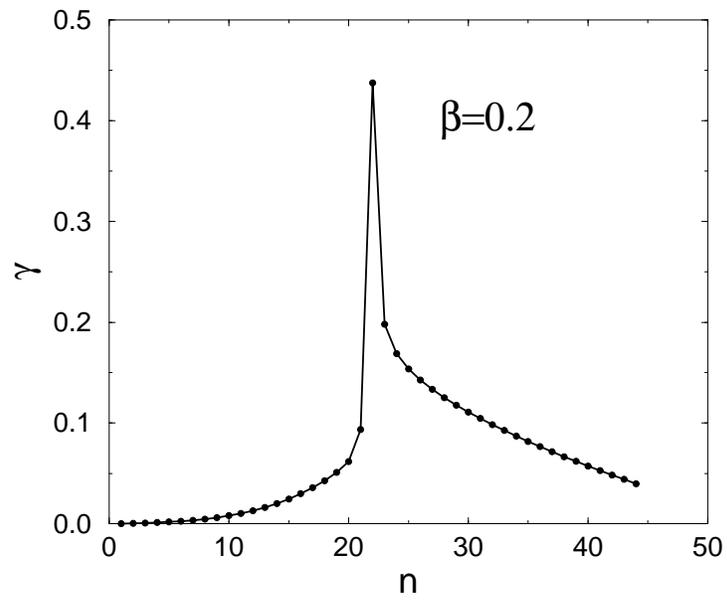
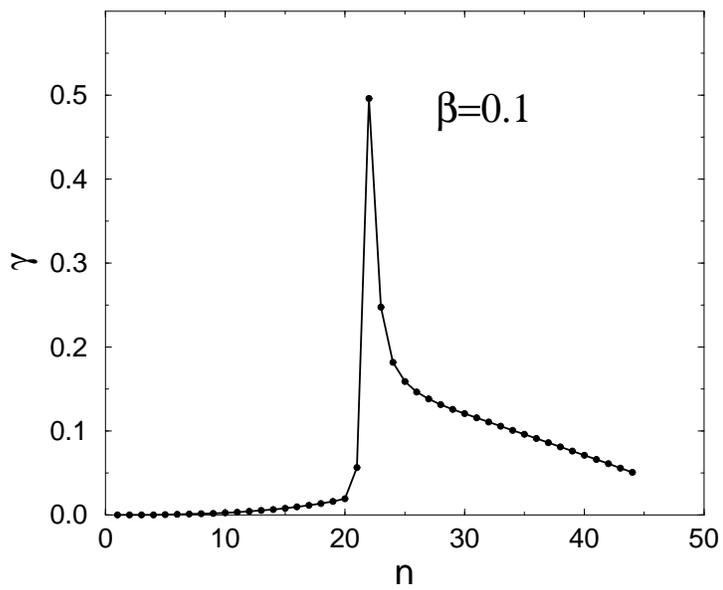
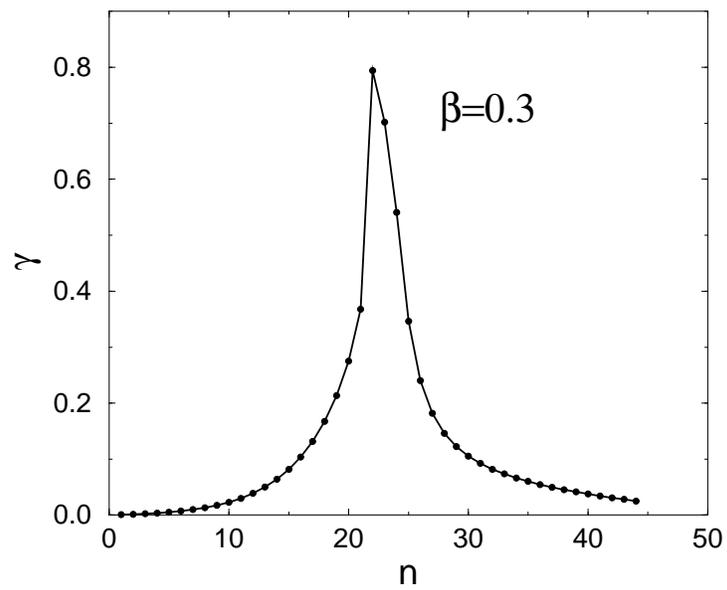
\end{document}